# Machines & Influence: An Information Systems Lens

Shashank Yadav[1]
IITB

**Abstract**

*Policymakers face a broader challenge of how to view AI capabilities today and where does society stand in terms of those capabilities. This paper surveys AI capabilities and tackles this very issue, exploring it in context of political security in digitally networked societies. We extend the ideas of Information Management to better understand contemporary AI systems as part of a larger and more complex information system. Comprehensively reviewing AI capabilities and contemporary man-machine interactions, we undertake conceptual development to suggest that better information management could allow states to more optimally offset the risks of AI enabled influence and better utilise the emerging capabilities which these systems have to offer to policymakers and political institutions across the world. Hopefully this long essay will actuate further debates and discussions over these ideas, and prove to be a useful contribution towards governing the future of AI.*

## 1. Introduction

All Information Systems are fundamentally a sociotechnical conception, comprising of networks of humans and their ideas, software, and hardware components which interact together to capture, process, generate, store, and deliver information (A. S. Lee, 2010). While the traditional notion of Information Systems was limited to the organisational interactions with computational technologies (Alter & Alter, 1996). However, lately the growth and developments in Artificial Intelligence (AI) have necessitated a more comprehensive approach to understanding the interaction between Information Systems and the whole of society, particularly to integrate present day AI capabilities into an Information Systems' perspective. Society being a very broad entity, we can focus our understanding of the man-machine interaction within the Networked Information Environment (NIE). Our conception of NIE expands upon the idea of Digital Communication Networks (Waghre & Ramprasad, 2021), and includes:

- All computer network based products and services irrespective of their business model, geography, demographics, price or lack of thereof
- The ecosystem of organisations and individuals which builds and operates the above
- Governments and regulatory bodies tasked with policing and enforcing laws on the internet, and the digital tools they use for that purpose
- The physical infrastructure which enables the internet and other computer networks
- All individuals who form the current or potential userbase for the above, or are associated with above, and their mutual interactions over the internet
- All types of content on the internet, including code, bots, and AI generated personas

---

[1] 214468003@iitb.ac.in / shashank@asatae.foundation

The complex and evolving information systems have always faced an everyday problem of loss of information, offsetting which had been the goal of the early ideas on information management (Berners-Lee, 1989). These ideas over the years evolved into the view of complex organisations entirely composed of building blocks of information systems, particularly computer and communication technologies (Jarvenpaa & Ives, 1994), where people simply coupled and decoupled from knowledge nodes. These ideas further evolved into the view of information management as necessary to storing, using, and exchanging credible information where elements of organisational consensus and contracts (M. Li et al., 2021) also came into the policies and management processes related to strategic and mission-critical information systems.

While the inception of AI had been a deep philosophical interjection in technology asking whether machines could think (Turing, 1950), since the creation of the connectionist paradigm of AI development which was described by its creator as a pattern recognition device (Rosenblatt, 1957); AI, and particularly Neural Networks, have become so increasingly data intensive to the extent that much of what constitutes operating AI systems can also be described in terms of information handling and management (O'Leary, 2013). Here, we are distinguishing information from data by understanding it as something contextual (Earley et al., 2017).

Taking into consideration the trend of dedicated software and hardware platforms (Gadiyar et al., 2018) for accelerating different AI workloads and the setting of the objectives and lifecycle goals of how these systems are used in different social and political contexts, it becomes clear that the existing socio-technical layer of Information Management has structurally penetrated the paradigm of machine intelligence as well. Infrastructurally, this layer has the basic software, computing, and network infrastructure of the data management workloads, but additionally it also has the power optimising hardwares embedded everywhere from computers and autonomous devices to data centre class accelerators (Reuther et al., 2021). The "human-in-the-loop" continuous capturing, managing, labelling and curating of information for building, training and operating all the algorithms, softwares and hardware components of AI applications – adds into the compounding complexity of a new and more contextual layer within the traditional data management paradigm.

Henceforth, we can understand this new layer of information management activity in the development and deployment of AI systems as a structural archetype for goal driven end-to-end cycles of transformations, storage and flows, of information to establish the functioning of machine intelligence. AI systems further reinforce this by autonomously capturing, processing, generating, and delivering information – as we will study in the subsequent sections.

Given that the fundamental idea of AI pertains to actuation based on sensor inputs (Russell et al., 2016), various AI techniques can be intuitively clustered into two contexts of of AI based information operations – one where AI is used to sense the information environment, and another where it shapes that environment. Therefore the next section, Section 2, covers Machine Sensing, i.e. capturing and processing information. Section 3 explores Machine Shaping, i.e. generating and delivering information. In section 4 we bring everything together in context of AI based political influence, introduce a Matrix of Machine Influence, and end with some concluding remarks.

## 2. Machine Sensing

### 2.1. Capturing Information

The history of AI was earlier dominated by the use of heuristics (Newell & Simon, 1976). There was little focus on capturing information about the real world, as compared to building a symbolic representation of it. This approach to AI development led to many excellent expert systems too, such as the Chess playing computer Deep Blue, which summarily defeated World Champion Gary Kasparov in 1997. Ironically, it was Gary who had predicted in 1988 that no AI will defeat a Grandmaster before 2000. The Deep Blue system was driven by faster hardware and the rules of Chess were embedded in the wiring of an eight-by-eight combinatorial array representing a real world Chess board over a chip (Hsu et al., 1995). The chip performed evaluation, search, and move generation. That is how early AI "captured information" about the world – the information was simply embedded into it.

This is in stark contrast to today's AI, which learns the rules of the game by playing it. This transformation in information capture was led by a roboticist who proclaimed that "the world is its own best model, always exactly up to date" (Brooks, 1990) and the engineers therefore just need to sense it more often and more appropriately. While his proclamation had nothing to do with how Neural Networks function, his ideas on grounding the machines in a reality that is "out there" heavily contributed to how later AI development embarked upon the evolution from symbol based construction to building and operating intelligent systems using real data about the world. Persistently seeking and managing that data about the real world is now the lifeforce of all modern AI-based information systems (Buxmann et al., 2021).

This extraction of information is essential for building the knowledge base which would serve as the AI's model of the world, and help the machine understand incoming data. There are two broad approaches to process incoming data, as a stream or as a batch. Stream processing is often referred to as online learning and batch processing as offline learning (Ben-David et al., 1997). These should not be confused with on-policy and off-policy learning in reinforcement learning based algorithms.

Online learning represents the entire family of machine learning methods (Hoi et al., 2018) where an agent (or model) attempts to solve a decision problem by learning from a sequence of data instances one at a time, like processing a literal stream of data. This is more suited for real-time operations. However, not all data is required immediately and when this stream of data instances are clubbed together into a batch, and fed to train the machine learning model while it isn't perpetually making parallel decisions, this can be referred to as offline training. Both these approaches bring in technical debts in form of unstable and underutilised data dependencies which cost more than code dependencies in any AI system (Sculley et al., 2015). Therefore also making it necessary that more organisational attention and resources are allocated to the pipelines of capturing information. Sensing, thus, and especially sensing more often and appropriately, takes over shaping in practitioners' priorities. For policymakers too, it entails that any attempt at the governance of AI has to begin with the governance of data and information pipelines.

To influence the NIE using AI technologies thus cannot be conceived without an adversary optimising and strategically (or tactically) deploying his AI-based sensing capabilities. An important aspect of this is how information and relations will be discovered from unlabelled data instances, also referred to as Open Information Extraction (Open IE). In contrast to mere Relation Extraction which requires relatively more human involvement in form of supervision and painstaking labelling (Jiang et al., 2020), the 2007 development of unsupervised Open IE (Banko et al., 2007) for extracting relational tuples from a corpus of 9,000,000 webpages without any human inputs brought a paradigm shift in information capture that continues today over not just text but all forms of unstructured data on the internet. Interestingly, nearly at the same time in an independent work, another set of researchers had come up with 'Unrestricted Relation Discovery' over a collection of newswire articles (Shinyama & Sekine, 2006).

While these can be considered early developments in capturing information from unstructured data, they set the tone and spirit of informational sensing in terms of "open and unrestricted extraction" from the internet, which greatly shaped future thinking and research.

An interesting subsequent development in capturing information happened when the term 'Neural Information Retrieval' appeared in 2016 during a Microsoft Research workshop in Italy. Most of the workshop proceedings were about whether and how Neural Networks can boost information retrieval ('Neu-IR', n.d.). Information retrieval differs from extraction in that it is a search problem where a query is made upon data, and the latter refers to structuring previously unstrucxtured information into a database (which can then be queried). In 2020, Facebook AI research brought Neural Networks to retrieve and extract information from unstructured data using natural language (Thorne et al., 2020), something we'll explore in the next section while discussing AI reasoning capabilities.

All this information retrieval and extraction leads us to the larger sensing goal of persistent ICB (Identity, Content, Behavior) profiling. Identity, content, and behavior become the key sensory signals which inform the inferences of feed algorithms at the user end of the information pipeline (Jurka et al., n.d.). In fact this measurement of social phenomena is unavoidable in sociotechnical systems (A. Z. Jacobs, 2021) and thus sociopolitical values are continuously encoded into the wider information system. This sociotechnical phenomena is already used for large scale response prediction where there is a machine learning model optimising discreet, short-term activity for each user (X. Zhang et al., 2016).

In political contexts, such micro-level sensing by an adversarial entity would equate to building adversarial capacity for subversion. Therefore for a political system struggling with AI-driven platforms and manipulative feeds, it becomes pertinent to understand the organisation of machine's sensory inputs. This capability to sense the NIE isn't useful only for an adversary but equally useful for the state to counter the adversarial influences. For example, the detection and classification of socio-political events, particularly understanding the dynamics of protest events (Hürriyetoğlu et al., 2021), is very useful for the state and its adversaries alike, as is the ability to sense and measure bias and polarity in a population.

Active work in techniques such as sentiment analysis, real-time machine learning (RTML) (Nishihara et al., 2017), knowledge graphs and semantic data discovery (Abu-Salih et al., 2021) allows an actor to develop a continuously evolving ontology of a country's politics. These extraction capabilities carry over into the subsequent active measures (Rid, 2020) type of operations. For example, persistent capture enables practitioners to turn their extracted knowledge of NIE into a graph based multidimensional ideological space (Monti et al., 2021) where social data traces and information propagation models can be used to learn the opinion dynamics (Monti et al., 2020) within a population. This capturing of the causality of political opinions, amongst other things, helps create interpretable ideological embeddings of a society and thus predicting unobserved behavior during information cascades (Monti et al., 2021).

While the problems discussed earlier, of data dependencies and pipeline jungles (Sculley et al., 2015), still persist for every actor, in the globalised arena of politics there is a differential of knowledge, skill, and resources, which favors those who can minimise those costs and sense the NIE "appropriately and often enough". The combination of strategic and tactical utility of sensing the NIE by persistently capturing information from it provides the political impetus for the conversations that guide technology policy, acquisition, as well as advocacy (Schneider et al., 2020) for cyberspace resources and requirements.

As we've seen, information retrieval and extraction have come to play a major role in building the knowledge-based sociotechnical foundations of Machine Influence. Let us now understand where AI capabilities are in processing the environment as humans do, and how that affects our information environment.

### 2.2. Processing Information

We've seen that AI technologies enable a unilateral information capture. However, given a lack of high level abstractions and metacognition, there can be said to be a qualitative difference (Bengio, 2014) between how AI systems and humans perceive the social environment. The non-affective precision, the scale and speed of operations, and the capacity to overwhelm human expertise in narrow domains makes the AI processing environment a strategic and tactical asset which can become greatly instrumental in achieving the social, political, or military goals of a state. In fact, it is this importance and criticality of data-intensive information systems for the modern state, which can often even turn their operators into security vulnerabilities (T. Thomas, 1998). With relation and information extraction from a corpus of human language text, we saw some aspects of how AI capabilities to process the online environment augment Machine Influence over NIE, we'll now shift our focus to visual information.

The beginning of computer vision wasn't rooted in attempts to mimic human vision, but was rather a humble attempt by Larry, a PhD student at MIT in 1960, at recognising a three-dimensional polyhedra from its two-dimensional view (Roberts, 1963). Interestingly, that PhD student gave up on computer vision, joined DARPA, and is now better known as one of the inventors of the TCP/IP based modern internet. However, perhaps the most conceptually influential work in the area has been David Marr's Vision (Marr, 2010), which was first published in 1982 after David's untimely death, where he approached human vision as a complex information processing system. Today, we have reached a place where scholars are attempting to explain the system of human vision using ideas from how AI learns (G. E. Hinton, 2010) to represent incoming visual information.

The progress has been very fast lately, especially after the application of Convolutional Neural Networks (CNNs) in 2012 led to striking performance increments over existing computer vision benchmarks (Krizhevsky et al., 2017). Because of CNNs, the accuracy rate of image recognition softwares has gone from a mediocre 70-80 percent to almost 100 percent today (D. Zhang et al., 2021). This has dramatically changed how information is monitored across the NIE. Actors can detect adversary actions within seconds with fast and accurate analysis of satellite streams and social media images. At the strategic level, the tactical applications of these developments allow states to improve their negotiation prospects, observe adversarial behavior and better assess the nature of the conflict (Vaynman, 2021).

Integrating these developments with developments in language processing, such as the development of Transformers (Vaswani et al., 2017) which allowed massive parallelisation with respect to how linguistic input was processed and translated, in turn enables actors to formulate and operationalise a more holistic and responsive perception management strategy on the internet. Their use can become much more politically sensitive (Liu et al., 2021) when seen through the lens of generative capabilities, which we'll explore subsequently in Section 3. The development of Transformers however did enable a more effective computational modelling of political communications (Huguet Cabot et al., 2020), especially of the elements of rhetoric, metaphors, and emotions which are so intrinsic to human politics.

Yet, the enhanced processing of language and visual information is only the tip of the iceberg of Machine Influence, which is further augmented greatly from another key dimension of AI which has been central to its historical development, the processing of reasoning tasks.

The capacity to reason about information after all is essential to the capacity to interact with humans in meaningful ways, or understand human affairs in consequential manner. Any interaction requires the ability to answer questions, and indeed Question Answering (QA) engines like Wolfram|Alpha (Cassel, 2016) have been at the forefront of integrating and retrieving complex information for open-domain QA, the backbone of which is reasoning (Asai et al., 2020). Such answering engines incorporate multidimensional information capture and natural language processing to compute an answer. Over the course of time, such reasoning engines given their accuracy, build greater trust with users, as was also evident when 90 percent of nurses with access to IBM Watson prefered to follow (Upbin, 2013) the guidelines of IBM Watson. It must be observed here that trust is an indispensable instrument of influence as it shapes behavioral outcomes (Bligh, 2017) and ensures willing compliance.

AI reasoning capabilities further shape the nature of human-machine interaction. For example, in Facebook's Neural Databases, the capabilities to reason with facts in natural language guide (Thorne et al., 2021) the model to generate the optimal answers. Such capabilities enable the model to find supportive evidence in large unstructured data, which is typically how information in the NIE is situated.

As we've seen in Section 2.1, heuristics and symbols had dominated earlier approaches to AI reasoning. However today, the game of Chess (Silver et al., 2018), and Go (Silver et al., 2016), can be considered as the key benchmarks of human reasoning which have been already overcome by Neural Network based approaches. Researchers have also noted how AI reasoning capabilities are positively correlated (Polu & Sutskever, 2020) with the size of the AI model, which is a complementing factor to the emerging trend towards large models whose mere scaling could result into new capabilities (Bommasani et al., 2021).

In fact, the costs associated with training and scaling AI models has prompted some researchers to successfully train their models over decentralised and unreliable hardware (Ryabinin & Gusev, 2020), a core feature of the NIE. There are, however, still many open questions in this area given the lack of publicly available specifications and standardisation of requirements(Ignatov et al., 2018), but accelerating neural networks over mobile chipsets is a very active area of contemporary AI research (Ignatov et al., 2019) where researchers are trying to bring deep learning to portable devices and smartphones.

The capability to learn over networked devices is remarkably crucial to reasoning over an Internet of Things (IoT) infrastructure (Firouzi et al., 2021), in fact distributed reasoning (Oren et al., 2009) and processing has been a key driver towards the information processing paradigm of bringing code to the data, and not data to the code. This brings to importance the ideas of knowledge distillation (G. Hinton et al., 2015) and model compression, which become not only complementary to distributed reasoning in a networked environment but also a practical necessity (T. Li et al., 2020) for the federated learning of AI where NIE forms the training infrastructure.

Scholars have also argued that the ability of AI to process reasoning related tasks further enhances the existing information capture capabilities (Asai et al., 2020) and makes the AI better at traversing and understanding the NIE.

## 3. Machine Shaping

### 3.1. Generating Information

Some researchers have called AI based information generation as "anti-creative" (Loi et al., 2020) owing to the possibility of harmful mutations and costs to society. However, the area of computational creation of information, once considered science-fiction, has been advancing so rapidly that the creators of GANs (Generative Adversarial Networks) who basically brought the field to life in 2014 (Goodfellow et al., 2014), termed it infeasible for any survey to effectively summarise what is the state-of-the-art in the area just a few years later (Goodfellow et al., 2020). Interestingly, Jürgen Schmidhuber had also developed a similar architecture over a couple of decades before this (Schmidhuber, 1992), however his idea was perhaps ahead of his time and computing potential. Today, Ian Goodfellow is known as de facto inventor of GANs.

While there have been other generative methods and many other variations of GANs (Radford et al., 2016) (Arjovsky et al., 2017) (Karras et al., 2018) as well for undertaking a range of novel generative tasks from image to image transformations (Isola et al., 2017) and text to image (H. Zhang et al., 2016) to even text to speech synthesis (Yang et al., 2017) etc – we will follow the advice of Mr. Goodfellow and not get into surveying the variations of generative AI models. Instead, we will explore how Generative AI capabilities in general present more severe strategic challenges than other AI methods because of their informationally offensive nature.

Researchers have successfully used GANs to produce false satellite imagery (Xu & Zhao, 2018) which forms a crucial aspect of how modern conflict narratives evolve. Open access satellite data has already led to reduced state capacity for subterfuge during geopolitical conflicts (The Economist, 2021), therefore the addition of generative AI capabilities holds a promise and a compensation for engaging in conflicts over the narratives of conflict that extend well over into the NIE. It is not just the creation of consumable content but any information, such as the network traffic and other technical characteristics of popular online platforms (Rigaki, 2018), a technique which can be transfered to tactically mimic the critical infrastructure and command-and-control centres' digital footprint, which puts information generation capabilities of AI in the centre of the constantly evolving puzzle of strategic security.

The recent development of Open AI's Codex (Chen et al., 2021) which lets users accurately generate and complete computer programs, forms an incipient indication of a future where the most challenging reasoning tasks which require a logical generation of information, could be turned over to the machine. To any actor with enough computing power and access to advanced models like these, this would mean the capability to develop novel software tools which tilt the balance of information dominance in cyberspace. But in terms of influence, there could be far deeper implications of the rise of generative models, such as the process of training these models could also generate in these models an understanding (Karpathy et al., 2016) of the world.

Another concerning and emerging reality is that the generative AI models based on Transformer architecture, like the GPT series, will be feeding into the rise of "no code platforms" and highly optimised computational generation of advertising content (Floridi & Chiriatti, 2020). Think-tanks have already noticed this and flagged the threat of automated autonomous disinformation using models like GPT-3 (Buchanan et al., 2021), arguing that these models are better suited for generating disinformation than humans because they enable an adversary to scale information operations across the NIE and try out far wider messaging options and variations.

Notwithstanding the contested uses, the capability to produce synthetic information has a great many upsides. One of these which concerns perhaps all actors on the internet is the organisational capability to generate synthetic data as an information anonymisation method (Arnold & Neunhoeffer, 2020). While some scholars have argued that generating synthetic data also brings the same challanges of balancing the privacy-utility tradeoff (Stadler et al., 2021), others have showed that further work in synthetic data could hold greater promises in differential privacy (McKenna et al., 2021). Researchers have also tried to solve the more pressing constraints of AI development, that of requiring a large corpus of labelled data for training or dealing with unseen output classes, by using computationally generated training data (Wang et al., 2021) and synthetic CNN features for classifying unseen classes (Xian et al., 2018). Famously, Tesla also uses synthetic information based simulations to train the autopilot (Tesla, 2021) of their entire fleet of interconnected autonomous vehicles which also learns in tandem from real-world data when out on the streets. It has also been found much useful to create massive datasets based on AI generated information where real-world data sources prove costly and infeasible (Shermeyer et al., 2021), for example, the imagery of aircrafts from an overhead perspective.

Between GANs and the older generative models, GAN based approaches have consistently shown better performance than other approaches in generating natural imagery (Y. Zeng et al., 2019), explaining the rise in both strategic interests and as well as concerns. Towards the goals of extending influence, generative AI should be seen as having a catalysing and pronouncing effect on the existing methods, tactics, and motivations. Just as technological enabling makes Ransomwares emerge as the digital extension of the human tendency to take something hostage for bargaining (Gilbert, 2021), the promises of generative AI find widespread adoption in the strategic sector owing to the age old political-military utility of deception (Holt, 2010) and "potemkin personas" in the cyberspace (DiResta & Grossman, 2019). Understandably therefore, scholars have described DeepFakes as the "newest way to commit the oldest crimes" (Spivak, 2019).

The utilisation of these technologies over the NIE brings to fore the traditional dilemmas of governing dual usage which depends to a great degree on the value system underlying the politics (Forge, 2010). Researchers have duly underlined that generative AI will enable a future of new forms of experiences and artistic expressions (Esling & Devis, 2020), also noting that while these may pose a new threat to national security, elections, and political stability - the world, its laws and policing, need to prepare for a world of synthetic information (C. Zeng & Olivera-Cintrón, 2019) which in form and function could be indistinguishable from the natural. The dual use quandary is inescapable, researchers have noted the GANs' use to strengthen a cybersecurity system (Kim et al., 2018), improve the quality of privacy and secrecy (Huang et al., 2018), and also to cause adversarial attacks on the very same ecosystem (Hu & Tan, 2017). This dual use nature is inherent in all technologies and sciences as Richard Feynman had noted much earlier (Feynman, 1955), which persistently affects the perceived modalities of deploying any machine intelligence with an objective of shaping the NIE.

### 3.2. Delivering Information

The abstraction of machines 'delivering information' does not refer to any messaging service or the communication protocols behind sending and receiving information, instead it refers to what people see when they go online. With greater social and economic activity shifting over to the NIE, there has been an ubiquitous operational transformation towards using Neural Network based approaches to finetune (Covington et al., 2016) the recommended information over large communication platforms. In such platforms, the AI system may have to juggle multiple objectives of ranking information. Here an older and popular technique of using a mixture of expert neural networks playing divide-and-conquer over the search space, which first appeared in 1991 (R. A. Jacobs et al., 1991), is still used with much success as Mixture-of-Experts based AI architectures are widely deployed (Zhao et al., 2019) to maximise both the content satisfaction and the user engagement objectives of large communications platforms.

However, the artificially intelligent delivery of information today takes its roots in the early need for automated selection and delivery of information (Foltz & Dumais, 1992) which had emerged in early 1990s as a result of increasing information overloads and the organisational incapacity (Rock, 2016) to manage it. Those early days of struggles with information management would seem archaic if we just take into account the context aware information distribution which was emerging within a decade (Marmasse & Schmandt, 2000) owing to the rise in mobile devices. Today, from military decision support systems (Weber & Aha, 2003) to on-demand education across the world (Konjengbam & Nagayoshi, 2021) rely heavily on machine intelligence based information delivery methods.

An interesting hypothesis has also emerged that algorithmically delivered consumption of information over time leads to consuming less diverse information (Spotify, 2020), an effect which may also compound political homophily in online communications (Bond & Sweitzer, 2018). Some researchers have also noted how dependence over AI based selection and delivery of information could also skew the delivery in a manner unintended by the organisation (Ali et al., 2019). That said, it is not merely the delivery of information but the entire planning process underpinning the organisational communication workflows which has been shifting over to AI based methods (Qin & Jiang, 2019) for discovering insights about the audience, generating optimal content, bidding and pricing for online spaces, and evaluating the outcomes at large scale for each target audience.

Scholars have noted that information systems can contribute significantly to decision variability (Elson et al., 1997), which also makes the pipeline of information delivery sensitive to adversarial attacks. And indeed, communications based platforms of all types face an everyday challange of intelligent content delivery (O'Hare & O'Grady, 2003) where increasing transmission security and decreasing delivery latency (Misra et al., 2013) emerge as their primary concerns in managing the logistics of delivering meaningful information. In fact as more and more connected critical devices like intelligent vehicles join the NIE, reliance on secure AI assisted information delivery (Sun et al., 2020) to enable myriad applications and services will only further increase.

Today, various communications inside the NIE can be broadly categorised as machine-to-machine, machine-to-human, human-to-machine, or human-to-machine-to-human. Perhaps, this overt predominance of machines and machine intelligence can also provide a glimmer of hope in the post-truth world by creating what some scholars have referred to as a "truth layer" (Westerlund, 2019), an automated AI-based content authentication layer integrated into the internet to provide some measure of the authenticity of the content being delivered. This is easier to say than implement as interoperability and maintenance issues will not be limited to any one component but span networks, databases, and even GUIs.

Information delivery after all is a two way street, like pulling and pushing (Klumpe et al., 2020). A Pull happens when information is delivered from the user side, willingly or not, and a Push happens when information is delivered to the user side, with or without user consent. We can see that information Pull is closely related to information capture we discussed in Section 2, as most of the context on which machine decisions are based is pulled from the user side, resulting in the highly context aware and localised nature of information being pushed back. Users' responses to the information delivery can then be measured (Gharibshah & Zhu, 2021) and pulled for finetuning the next Push based on the predictive response behavior.

AI driven delivery of information over large communication platforms also runs into a myriad set of issues. Chief among these are filter bubbles, echo chambers, and algorithmic biases. Filter bubble (Pariser, 2011) can be thought of as a selective exposure to information because of a search based circularity of perspectives which is also deeply intertwined with users' own personalised preferences and activities. Echo chamber is a more general term for describing the process and outcomes of political homophily in NIE (Petrov & Proncheva, 2018), scholars have noted how this phenomena leads to strategic voting behavior which affects the electoral outcomes by skewing the social network structure towards bipolarity (Tsang & Larson, 2016). Lastly, while there is also the fact that not all biases are bad and the emergence of bias as a phenomena is inherent in the process of recognising patterns in groups (Danks & London, 2017), the more politically contentious biases in AI can also result more from the poor data handling processes within the organisational machine learning pipelines (Bozdag, 2013).

All of the above information delivery issues are also social and political liabilities which can easily become an instrument in the toolbox of an adversary in shaping the NIE by misaligning the autonomous information delivery with the larger sociopolitical objectives of the defender.

## 4. Discussion & Conclusion

Considering the vastness and diversity of our information environment, it is prudent to contextualise the political threat of AI applications using a matrix based on the nature of AI usage, i.e. sensing and shaping information. To undestand AI capabilities in the operational contexts of information influence, we further incorporate the dimensions of the strategic and tactical aspects which are integral to all political-military influence operations (Clausewitz, 1989) and also applicable to information operations in our Networked Information Environment. To make a quick note - strategy is generally understood as a plan to achieve the larger political objectives, and tactics as the operational methods to execute that plan - strategic planning thus rests on tactical success. Strategic behaviour is therefore more high-level, continuous and less subject to change. While tactical behaviour on the other hand is quite the opposite – ground level, discontinuous, and subject to much improvisation.

We thus produce the following, elementary, 'Matrix of Machine Influence' (MMI):

|  | **Machine Sensing** | **Machine Shaping** |
|---|---|---|
| **Strategic** | *Strategic AI Sensing* *[Continuous]* | *Strategic AI Shaping* *[Long-termist]* |
| **Tactical** | *Tactical AI Sensing* *[Discontinuous]* | *Tactical AI Shaping* *[Interventionist]* |

The key to exploring the MMI is understanding that the political actions happen in a continuum of a policy feedback loop from strategy to tactics and vice versa (Handel, 1992), where all operations are under a constant stress from strategic and tactical considerations both. Each of the boxes of our matrix can be turned into an N-player matrix game to evaluate the best policy under specific circumstances. For example, we can see with just a very basic 2-player game between symmetrical players that pervasive and continuous sensing when another player is adversarial, is an optimal information policy for both players given the risks of information asymmetry.

Here, $B_1, C_1$ : Benefits, Costs for player one; and $B_2, C_2$ : Benefits, Costs for player two; where $B > 0 > C$

|  | Player 2 ⇩ | |
|---|---|---|
| Player 1 ⇩ | **Sensing** | **Not Sensing** |
| **Sensing** | *B1, B2* | *B1, C2* |
| **Not Sensing** | *C1, B2* | *0, 0* |

Or that in the above game itself, under the additional circumstances of adversarial shaping from the other player, a policy of inactivity will be costly, and if both players engage in adversarial shaping, both will suffer appropriate costs.

|  | Player 2 ⇩ | |
|---|---|---|
| Player 1 ⇩ | **Sensing** | **++ Shaping** |
| **Sensing** | *B1, B2* | *B1-C1, B2+B2* |
| **++ Shaping** | *B1+B1, B2-C2* | *B1-C1, B2-C2* |

These are but highly simplistic representations above, to show that further work in calculating complex payoffs for each AI operational vertical (information capturing, processing, generating, and delivering) under different strategic and tactical considerations, integrating time constraints, risk aversion, policy combinations and multiple N-player profile settings etc can yield a much more comprehensive and useful tool for policymakers to navigate the machine driven political landscape of the NIE. All that is however, beyond the scope of this work.

It is noteworthy that 'the machines' as an abstraction have been described (McLuhan, 1994) as fragmentary, centralist and superficial in the patterning of human relationships where like the bee of the plant world, man becomes the sex organs of the machine world, enabling the machines to "fecundate and evolve ever new forms". Irrespective of which side of the debate one is on, the fact remains that machine intelligence has deeply penetrated the functioning of modern human society, its networked economies and everyday politics. This long essay has provided sufficient literature to make such a case. Perhaps this intertwining of man and machine raises more than just debates, as some radical elements (Kaczynski, 1995) have even gone all out against any intelligent machines altogether. We have, however, taken a substantially different position from this pessimism and suggested better information systems management/regulation to more optimally offset the risks and utilise the emerging capabilities which AI systems have to offer to policymakers and political institutions across the world.

Towards the above objective, in this paper we explored the socio-technical AI capabilities in the terms of information management, where the capturing, processing, generating, and delivering of information were established as the capabilities which enable actors to influence, i.e. sense and shape the NIE and possibly shift the balance of political stability and security. To navigate the AI enabled socio-political influences, we introduced a Matrix of Machine Influence to assist the practitioners and policymakers in evaluating their policies and actions. Further, we suggest that future interdisciplinary research in integrating these ideas from a game theoretic perspective could be much useful in governing the harmful applications of machine intelligence, and in ensuring political security within digital societies.

## Acknowledgement

I am deeply grateful to Dr. Sundeep Oberoi for guiding the trajectory of this work and providing his support and inputs throughout the process.

***